\documentstyle[12pt,epsf]{article}

\date{\today}

\newcommand\putfig[3]{
   \vbox{
   \let\picnaturalsize=N
   \def\picsize{#3}
   \def\picfilename{#1}
   \ifx\nopictures Y\else{\ifx\epsfloaded Y\else\input epsf \fi
   \let\epsfloaded=Y
   \centerline{\ifx\picnaturalsize N\epsfxsize \picsize\fi
   \epsfbox{\picfilename}}}\fi
   \vspace{1.0cm}
   {\it #2}
   \vspace{1.5cm}
   }
}

\def\be{\begin{equation}}
\def\ee{\end{equation}}
\def\bear{\begin{eqnarray}}
\def\eear{\end{eqnarray}}
\def\nn{\nonumber}

\def\ket{{\rangle}}
\def\hlf{{{1\over 2}}}

\def\const{{\mbox{const\ }}}                    

\def\lbr{{\lbrack}}
\def\rbr{{\rbrack}}

\def\wdg{{\wedge}}                              % wedge product

\newcommand\px[1]{{\partial_{#1}}}

\newcommand\inv[1]{{1\over{#1}}}

\newcommand\rep[1]{{\underline{\bf {#1}}}}      % representation
\newcommand\MS[1]{{{\bf S}^{#1}}}               % Circle, sphere,...
\newcommand\MT[1]{{{\bf T}^{#1}}}               % Torus
\newcommand\CP[1]{{{\bf CP}^{#1}}}              % CP
\def\a{{\alpha}}

\def\u{{\mu}}
\def\v{{\nu}}

\def\s{{\sigma}}

\def\lam{{\lambda}}

\newcommand\ep[1]{{\epsilon_{#1}}}              % anti-symmetric tensor
\def\IC{{\bf C}}                                % Math Z
\def\IZ{{\bf Z}}                                % Math Z
\def\wdg{{\wedge}}                              % ^
\newcommand\kbl[1]{{{\cal K}_{#1}}}             % canonical bundle

\newcommand\degr[1]{{{\mbox{deg}}({#1})}}        % degree
\newcommand\SUSY[1]{{{\cal N}= {#1}}}           % N=? SUSY

\def\SF{{{\cal S}}}                             % Surface
\def\npb#1#2#3{{\it Nucl.\ Phys.} {\bf B#1} (19#2) #3}

\def\prd#1#2#3{{\it Phys.\ Rev.} {\bf D#1} (19#2) #3}

\def\hepth#1{{\it hep-th/{#1}}}

\begin{document}
\begin{titlepage}
\titlepage
\rightline{PUPT-1631}
\rightline{hep-th/9607020}
\rightline{July 2, 1996}
\vskip 1cm
\centerline {{\Large \bf A Test Of The Chiral $E_8$ Current Algebra}}
\centerline {{\Large \bf On A 6D Non-Critical String}}
                 
\vskip 1cm
\centerline {Ori J. Ganor
\footnote{Research supported by a Robert H. Dicke fellowship and
by DOE grant DE-FG02-91ER40671.}}
\vskip 0.5cm
\begin{center}
\em  origa@puhep1.princeton.edu\\
Department of Physics, Jadwin Hall, Princeton University\\
Princeton, NJ 08544, U. S. A.
\end{center}
\vskip 1cm
\abstract{
Compactifying the $E_8$ non-critical string in 6D
down to 5D the 6D strings give rise to particles and
strings in 5D. Using the dual M-theory description
compactified on an elliptically fibered Calabi-Yau
we compare some of the 5D BPS states to what one expects
from non-critical strings with an $E_8$ chiral current algebra.
The $E_8$ multiplets of particle states comprise of
2-branes wrapping on irreducible curves together with bound
states of several 2-branes.
}
\end{titlepage}

%%%%%%%%%%%%%%%%%%%%%%%%%%%%%%%%%%%%%%%%%%%%%%%%%%%%%%%%%%%%%%%%%%%%%%%%%%%%
%                      BEGINNING OF TEXT                                   %
%%%%%%%%%%%%%%%%%%%%%%%%%%%%%%%%%%%%%%%%%%%%%%%%%%%%%%%%%%%%%%%%%%%%%%%%%%%%

%==================================================================%
% Section (1): Introduction                                        %
%==================================================================%
\section{Introduction}

It was argued in \cite{GanHan} that the non-critical strings related
to small $E_8$ instantons carry a chiral $E_8$ current algebra.
In \cite{SWCSD,WitPMF} this algebra was realized as fermions
in the F-theory setting and in \cite{MVII} the $E_8$ lattice was given
a geometrical interpretation as the integral cohomolgy lattice of
a 4-manifold whose collapse is responsible for the phase transition.
The authors of \cite{MVII} suggested a way to identify geometrically
the states corresponding to this current algebra and the purpose
of the present paper is to work out some of the details of their
suggestion.

Compactifying the 6D theory at the phase where a non-critical string
appears down to 5D there is a dual description of M-theory on
a Calabi-Yau \cite{AFT,FKM,WitPMF}. The BPS string states in 6D become
stringy states and particle states in 5D and are identified with
2-branes wrapped on 2-cycles and a 5-brane wrapped on a 4-cycle
respectively \cite{WitPMF,MVII}.
From the non-critical string point of view, the states must fall
into representations of $E_8$ and the multiplicity
of particle states must correspond to the multiplicity in a chiral
$E_8$ current algebra \cite{MVII}.

In the situation at hand one has to consider bound states of 2-branes
wrapping on different 2-cycles that intersect \cite{SenUDU}.
Indeed, such states are needed to complete $E_8$ multiplets.
The general techniques for studying such bound states have been
developed in \cite{WitBST} and this particular problem of bound states
of 2-branes has been studied in \cite{SenUDU} where their existence was
necessary to complete U-duality multiplets.
An analogous question for 5-branes has been recently studied in
\cite{HanKle} where, remarkably, a non-critical string in 4D appeared.

Only slight modifications
are needed for our case -- first there is less supersymmetry
and the 2-brane motion is frozen in 2 dimensions (orthogonal
to the collapsing 4-cycle) and second we are in M-theory and not
type-II string theory.

Taking into account the bound states -- or {\em reducible curves} --
and using the techniques of \cite{VafIOD,BSVtop} and \cite{WitPMF}
for calculating the states that arise from the moduli space of
such curves and their Lorenz, $E_8$  and other quantum numbers
we can compare the geometrical computation with the expected
physical result from the low-energy of the non-critical string, i.e.
the $E_8$ chiral current algebra and transverse oscillations.
This comparison is similar to the identification of
the BPS states of the heterotic string with cohomologies
of symmetric products of K3-s made in \cite{VafIOD,BSVtop}.

The paper is organized as follows.
Section (2) is a review of some of the facts about
non-critical strings and the F-theory
construction as well as the compactification to 5D. Section (3)
is a geometrical calculation of the multiplicity of stringy states
in 5D. Section (4) deals geometrically with some of
the particle 5D states. Section (5) is devoted to 2-branes wrapped
on reducible curves, or bound states of 2-branes, and finally
in section (6) the geometrical results are compared to the physical
expectation. Appendix (A) discusses in more detail the $H^2(\IZ)$
cohomology of the almost del Pezzo surface that shrinks at the phase
transition \cite{MVII}. Appendix (B) includes some algebraic
details of the counting of curves.

I have recently learned of another paper which contains similar
results using different arguments \cite{KMV}.

%==================================================================%
% Section (2): E_8 non-critical strings in F-theory                %
%==================================================================%
\section{$E_8$ non-critical strings in F-theory}

This section is a review of some of the relevant facts from
\cite{DMW,MVI,GanHan,SWCSD,WitPMF,MVII}.
The purpose of this review is also to set the geometrical notation.

The 6D heterotic string vacua compactified on K3 where
described in \cite{MVI,MVII} as F-theory on an elliptic 3-fold.
In this description phase transitions at points where  tensionless
strings appear correspond to collapsed cycles in the 3-fold
\cite{MVI,SWCSD,WitPMF,MVII}.
In this paper, we will be interested specifically in vacua
near the point where a small $E_8$ instanton collapses.
In the physical language this point connects two phases \cite{GanHan,SWCSD}.
One phase is the finite size instanton which breaks
the $E_8$ gauge symmetry while the other phase has an extra 
tensor multiplet \cite{DMW} but the $E_8$ symmetry is restored (locally).
In F-theory this phase transition corresponds to blowing up a point
in the base of the elliptic fibration.

The geometrical setting that describes that phase in F-theory
is as follows \cite{WitPMF,MVII}. 

It is sufficient to restrict to the vicinity of the blown-up point.
The 3-fold on which F-theory is compactified has a 4 dimensional base $B$.
Picking local analytic coordinates $z_1, z_2$ on $B$ such that the blown-up
point is at $z_1 = z_2 = 0$ -- the blow-up replaces the single point
at the origin with an entire $\CP{1}$ such that when we approach the origin
along a line with fixed $\lam = z_1 / z_2$ with $z_1, z_2\rightarrow 0$
we land on the point corresponding to $\lam$ on $\CP{1}$.
It is allowed to blow-up a point if we can extend the elliptic fibration
to the $\CP{1}$ (the ``exceptional divisor'').
Writing the elliptic fibration as \cite{VafaFT}:
\be
y^2 = x^3 - f(z_1, z_2) x - g(z_1, z_2)
\ee
the Calabi-Yau constant holomorphic 3-form is given by
\be
\Omega = dz_1\wdg dz_2 \wdg {{dx}\over {y}}
\ee
In the vicinity of $z_1 = z_2 = 0$ the good coordinates are
$z_2$ and $\lam$ (away from $\lam = \infty$) or $z_1$ and $\inv{\lam}$
(away from $\lam = 0$).
The coordinates $x$ and $y$ are singular, but a change of coordinates
to $\xi,\eta$ where:
\be
x = z_2^{2n}\xi,\qquad y = z_2^{3n}\eta \label{xtoxi}
\ee
will be good provided that $f$ is of degree $4n$ and $g$ of degree
$6n$. In that case we have
\be
\eta^2 = \xi^3 - f_{4n}(\lam, 1)\xi - g_{6n}(\lam, 1)
\ee
The holomorphic 3-form becomes
\be
\Omega\sim - z_2^{1-n} d\lam\wdg dz_2\wdg {{d\xi}\over {\eta}}
\ee
so for $n=1$ $\Omega$ is nonzero and nonsingular on the exceptional
divisor. This fixes $f$ to be homogeneous of degree 4 and $g$
to be homogeneous of degree 6.
In terms of the number of 7-branes (i.e. singular fibers) we find
that the discriminant
\be
\Delta_{12}(z_1, z_2) = {{f^3}\over {27}} - {{g^2}\over {4}}
\ee
has 12 roots in $\lam = z_1/z_2$ on the $\CP{1}$ so 12 7-branes
pierce the exceptional divisor \cite{SWCSD,WitPMF}. Only a maximum
of eight of them can be given a perturbative type-IIB description 
simultaneously.

The resulting surface (2 complex dimensions) given by
The elliptic fibration over the exceptional divisor $\CP{1}$:
\be
y^2 = x^3 - f_4(\lam) x - g_6(\lam)
\ee
is the surface that collapses at the tensionless string point.
We will denote this surface by $\SF$.
Its only non-vanishing Hodge number (other that $h^{0,0}$ and $h^{2,2}$)
is $h^{1,1} = 10$ and its intersection form is
\be
E_8 \oplus
\left( \begin{array}{cc} 0 & 1 \\ 1 & 0 \\ \end{array} \right).
\ee
It was explained in \cite{MVII} that the $E_8$ is related to 
the physical $E_8$ gauge group of the small instanton and we will
explore this relation further in the next sections.

The 2-cycles are (as described in \cite{MVII}) given by
$f,b,e_1,\dots, e_8$, where $f$ is the fiber, $b$ the
base and $e_i$ are mixed. A typical 2-cycle that
is a combination of the $e_i$-s is given by picking
two 7-brane points (that is, points where
the fiber degenerates) $\lam_1$ and $\lam_2$ on $\CP{1}$
and a path $\gamma$ on $\CP{1}$ from $\lam_1$ to $\lam_2$.
At $\lam_1$ there is a unique   1-cycle  of the fiber
that shrinks to zero. As we go along $\gamma$ this cycle
grows. As we reach $\lam_2$ a 1-cycle has to shrink again,
if it is this cycle that we transported, then we have
drawn a 2-cycle which is of genus zero.
As this construction will be important in what follows,
we have described it in more detail in appendix (A).

Physically, the 6D F-theory vacua were described \cite{VafaFT}
as type-IIB string vacua on a 4-manifold (the base $B$ of the Calabi-Yau)
with 7-branes of different types filling the entire uncompactified
6D as well as analytic 2-cycles inside the base $B$.
When a point of $B$ is blown-up there is an additional small 2-cycle
from that point. The 3-brane of type-IIB can wrap on that 2-cycle
and this is the string with small tension that corresponds to the
transition \cite{SWCSD,WitPMF,MVII}.

In this paper we will concentrate on the small $E_8$ instanton
transition. The F-theory description shows that there are other transitions
in the moduli space corresponding to other 4-cycles that shrink \cite{MVII}.
Those (as well as the small $E_8$ have been recently discussed
in \cite{CPR}.

\subsection{Compactification to 5D}

Compactifying the six-dimensional theory on $\MS{1}$ with radius $R$
we end up at a point on the moduli space of 
the heterotic string on $K3\times \MS{1}$. 
That theory is equivalent to M-theory on a Calabi-Yau as discussed
in \cite{CCDF,AFT,FKM,WitPMF}.

We are interested in that region of moduli space that emanates
from the 6D small $E_8$ instanton point.
After compactification to 5D the $E_8$ heterotic string is dual
to $SO(32)$ but if we do not turn on $E_8$ Wilson lines on $\MS{1}$
the T-dual radius will be $R' = {{R}\over {1+2R^2}}$ \cite{GinTOR}
and will be never large. With a special value of the Wilson loop
we can reach the large radius of the $SO(32)$ theory and the
non-perturbative phase that emanates from the small instanton point
becomes the Coulomb phase of Witten's field theory for the 
small $SO(32)$-instanton\cite{WitSML}, i.e. a phase where the 
non-perturbative
$Sp(1)$ of \cite{WitSML} has a Wilson loop along $\MS{1}$.

Turning on an $E_8$ Wilson loop is equivalent to turning on certain
$C_3$ (the 3-form of M-theory) expectation values.

Finally, a further compactification to 4D brings us to the duality
between type-IIA on a Calabi-Yau and the Heterotic string on $K3\times\MT{2}$
first found in \cite{KacVaf}.
An extensive description of the various phases as seen in 4D has been
recently given in \cite{LSTY}.

\subsection{Counting states in 5D}

The strings of 6D can either wrap around the 6th dimension to give
a particle in 5D, or  be reduced to give a string state.
The 6D strings have an $E_8$ chiral current algebra on them \cite{GanHan}.
It thus is interesting to see how the $E_8$ chiral
current algebra is exhibited on the stringy states
and how the particle states fall into multiplets of $E_8$.

We will work in the setting of M-theory compactified on the elliptically
fibered Calabi-Yau.

The structure of BPS charges was described in \cite{CCDF,AFT,FKM}.
There are $h^{1,1}-1$ vector multiplets coming from K\"ahler 
deformations of the CY  except for the overall volume which
is in a hyper-multiplet.
The vector-moduli space is completely determined by the 
integer intersection numbers of $(1,1)$ cycles of the CY.
In 5D BPS charges are real and are determined from the intersection
numbers \cite{AFT}.

The M-theory low-energy description is applicable when
the size of the CY is large. On the other hand the area of the
fiber is $R_E^{-4/3}$ where $R_E$ is the radius of $\MS{1}$
in Einstein units.
Since the overall volume is in a hyper-multiplet
and decouples from the vector multiplets  we can extrapolate to
small overall volumes.

%==================================================================%
% Section (3): Stringy states in 5D                                %
%==================================================================%
\section{Stringy states in 5D}

The six-dimensional ``non-critical'' strings at the small $E_8$
instanton point have an $E_8$ chiral current algebra on them.

In F-theory language, the $E_8$ is realized as 16 left moving
fermions on the world-sheet as follows \cite{SWCSD,WitPMF}.
A D-brane analysis shows that on the intersection of a 7-brane
with a 3-brane there live 2 chiral fermions.
In our case there are 12 such intersection points so it would
seem that we have 24 fermions. The suppression of 8 fermion
zero modes out of the total 24 is similar to the suppression of
4 $U(1)$-s out of the 24 in the case of F-theory on K3 \cite{VafaFT}.
The situation in the latter case 
was clarified in physical terms in \cite{DougLi}
where it was shown that the interactions of these $U(1)$-s with
the 2-form fields in the bulk make some of the gauge fields massive.
The basic reason is that the 7-branes are not of the same $(p,q)$
type.

In our case, the suppression of 8 fermionic modes occurs because
of the $U(1)$ gauge field that lives on the 3-brane that wraps on
the base. This gauge field has no resulting dynamical degrees of freedom
but the fermions at the intersections of the 7-branes and the 3-brane
are {\em charged} under that $U(1)$. Because of the varying coupling
constant of F-theory and because we have to use S-duality to get
a local perturbative D-brane description when moving along the base, some
of the fermions are magnetically charged and some are electrically charged
in any global description.
The surviving fermionic modes are those that correspond to a global
$U(1)$ field configuration with sources at some of the 7-branes
(where fermion modes are excited).
Such $U(1)$ fields are linear combinations of the following configurations.
Take any pair of 7-branes and a path connecting them in such a way
that as we move one 7-brane along the path towards the other 7-brane
we end-up with {\em mutually local} 7-branes (i.e. 7-branes of
same $(p,q)$-type). Now we can take one 7-brane to be a positive source
of $U(1)$ and the other to be a negative source of $U(1)$ and as we deform
the 7-branes back to the original position along that path we used above
we end up with a global $U(1)$ configuration that is consistent
with the S-dualities that we have to perform to relate different patches
of the base.

The reason for describing the $U(1)$ field configurations in that way
is to make contact later on with the geometry of the base.
Indeed, the above description is in 1-to-1 correspondence with
the construction of the $e_1,\dots,e_8$ $(1,1)$ 2-cycles in $\SF$
discussed briefly in the previous section and elaborated in
appendix (A). In any case, the analogy shows that there are precisely
8 linearly independent such configurations corresponding to the $2\times 8$
massless fermion modes that generate the $E_8$ chiral current algebra.

After compactification to 5D the connection with the geometrical
description is even more manifest.
The stringy states come from 5-branes of M-theory which wrap around
the 4-cycle $\SF$ in the Calabi-Yau which is now physical.
On the 5-brane there lives a tensor multiplet and the zero modes 
of the anti-self-dual 2-form $B_{\u\v}^{(-)}$ will give rise to fields
on the world-sheet. The zero-modes which are proportional to the
cohomology classes (dual to) $e_1,\dots, e_8$ become 8 left-moving
chiral bosons on the string world-sheet. 
The bosons live on the $E_8$ lattice and thus correspond in a 
natural way to the chiral $E_8$ current algebra.
The remaining modes of $B_{\u\v}^{(-)}$ and the scalars
of the tensor multiplet are in 1-to-1 correspondence with transverse 
oscillations of the string.

\subsection{A note on the geometrical interpretation of the 6D fermions}

As an aside, let us give another geometrical interpretation for
the fermions on the intersection of a 3-brane and a 7-brane.
T-duality replaces the configuration with a 6-brane and a 
2-brane of type-IIA.
The reason we work with that configuration is that the 6-brane,
being a source for the 1-form which corresponds to
rotations in the 11th dimension (i.e. a Kaluza-Klein monopole),
is completely geometrical
and is described by having the $\MS{1}$ of the 11th dimension
in a non-trivial $U(1)$-bundle around the 6-brane.

Let the 6-brane be in $x_7=x_8=x_9 = 0$ and
let the 2-brane be in $x_1 = \cdots = x_6 =0$, $x_9 = a$
($x_0$ being time).
The scalar fields that live on the 2-brane correspond to translations
in the ambient directions \cite{WitBST}.
There is also a gauge field $A_\u$ which in 3D is dual to a scalar
$\phi_{11}$. This scalar corresponds to the coordinate of the 2-brane
along the 11th dimension which is a circle.
As we increase $a$ from negative values to positive values,
we make the 2-brane pass through the 6-brane.
This passage is accompanied by a twist to the $U(1)$-bundle
which is the compactified 11th direction.
Thus as the 2-brane emerges for positive $a$, the $U(1)$
bundle on it is twisted.
This means that as $a$ passes zero, we get an increase of
$2\pi$ in:
\be
\oint_C \px{j}\phi_{11} dx^j
\ee
where $C$ is a contour that surrounds the origin in 
the 2-brane two-dimensional space.
Now, $\phi_{11}$ is dual to the gauge field so:
\be
\px{j}\phi_{11} = \ep{jk} E_k
\ee
where $E_k = \dot{A}_k$ is the electric field-strength.
We find that there is a jump in $\nabla\cdot E$ in the origin,
which corresponds to a jump of one unit in the electric charge
at the origin.

Let us see to what this corresponds physically after T-duality.
If we wrap the $x_9$ direction around a circle $\MS{1}$,
and T-dualize we find a 7-brane in the $x_7 = x_8 = 0$
position and a 3-brane in the $x_1 = \cdots = x_6 =0$ position.
They both wrap around $x_9$ and the variable $a$ above
transforms into the $U(1)$ Wilson loop around $x_9$.
The statement about changing $a$ from negative values to
positive values is the geometrical manifestation of the fact
that the massless fermions that
live on the $7\perp 3$ intersection are charged.
When there is a Wilson loop, the modes around the $x_9$
are fractional $\psi_{n+a}$. When $a$ moves from a 
negative to a positive value, one mode from the Dirac sea
becomes of positive energy, so there is an effective
change of one unit of charge in the vacuum.

%==================================================================%
% Section (4): Particle states                                     %
%==================================================================%
\section{Particle states}

To count particle states we need to wrap the 2-brane
around analytic 2-cycles.
Let us first review the quantum numbers of such states.
The spin of the states was determined in \cite{WitPMF} as follows.
If ${\cal M}$ is the moduli space of such analytic 2-cycles
(including flat gauge connections of the 2-brane which will
be elaborated bellow) then the states correspond to the cohomology
$\bigoplus_{p,q}H^{p,q}({\cal M})$ and the  
representation in the little group $SO(4) = SU(2)_1\times SU(2)_2$
was determined in \cite{WitPMF} by performing a Lefschez decomposition
of $H^*({\cal M})$ into representations of $SU(2)_1$ (trivial under
$SU(2)_2$)
such that $J_3 = \hlf(p+q-{\mbox{dim}}_\IC{\cal M})$ and 
tensoring with half a hyper-multiplet
 $2(\rep{1},\rep{1})\oplus(\rep{1},\rep{2})$.
In particular, all those states have $SU(2)_2$ spins at most $\hlf$
and $SU(2)_1$ spins at most $\hlf{\mbox{dim}}_\IC{\cal M}$.

From the ``physical'' point of view, these states come from wrapping
modes of the non-critical string. These strings carry a left-moving
$E_8$ chiral current algebra at level $k=1$ as well as 4 free
fields corresponding to oscillations in transverse directions
and their right moving super-partners which are 2 copies of right
moving fermions in the $\rep{2}$ of $SU(2)_1$ (see e.g. the appendix of
\cite{GanHan}).

Scalar BPS states correspond to excitations of left-movers only 
\cite{SenUDU,VafIOD,BSVtop}. 

The BPS states have charges corresponding to the momentum
along the 6th direction, the ``winding number'' of the non-critical string
and the $E_8$ representation.
The 6th momentum is given by $L_0 - \bar{L}_0$ of the CFT and
we will soon identify it geometrically.

 In this section we will describe the particle states mainly in geometrical
terms. In addition to irreducible curves we will see that there are also
reducible curves or {\em bound states} of 2-branes.
 It turns out that they are indeed needed to complete $E_8$ multiplets
and to reproduce the expected results from the chiral current algebra
calculation. We will discuss the reducible curves more
in section (5) and perform a more detailed comparison with the
non-critical string states in section (6).

\subsection{Curves in $\SF$}

The cohomology class of a curve $D$ in $\SF$ is a combination
\be
D = n b + m f + \sum_{i=1}^8  l_i e_i
\label{dclass}
\ee
where in the notation of section (2), $b$ is the section
of the fibration, $f$ is the fiber
\be
b = \{ x = y = \infty\},\qquad f = \{z = \const\}
\ee
and $e_i$ are the the other 8 cycles with intersection form $E_8$.
The other non-zero intersection numbers are
\be
b\cdot b = -1,\qquad b\cdot f = 1
\ee
(the first equation is deduced from (\ref{xtoxi})).

The canonical bundle $\kbl{\SF}$ has first Chern class
\be
c_1(\kbl{\SF}) = -f
\ee
and the genus of the curve $D$ is given by
\be
2g - 2 = D\cdot D + D\cdot \kbl{\SF} = -n^2 + (2m-1)n + \vec{l}^2.
\ee

As explained in \cite{MVII}, the Weyl group of $E_8$ acts
on the divisors $D$. Performing a closed loop in the
moduli space (of parameters of $f_4$ and $g_6$) the surface $\SF$
returns to itself up to an automorphism which by the Lefschez
theorem acts as a Weyl reflection on the cohomology classes $e_i$.

 From formulas in \cite{AFT} we can see that $P_6$, the momentum along
the 6th direction $\MS{1}$, is given by
the coefficient  of the fiber -- $m$.
\be
P_6  = m. \label{momentum}
\ee

 Finally, the winding number of the non-critical string is $n$.

\subsection{single covers}

Single covers are states with $D\cdot f = n = 1$.
They are given by sections of the elliptic fibration.
We thus need meromorphic functions $x(\lam)$ and $y(\lam)$
that satisfy
\be
y(\lam)^2 = x(\lam)^3 - f_4(\lam)x - g_6(\lam).
\label{cubic}
\ee
Let us first take $x$ and $y$ to be polynomials.

If $x$ is a polynomial of degree $2d$ then $y$ is of degree $3d$.
The intersection $D\cdot b$ is given by the intersection with
the generic section $x=y=\infty$.
$x(\lam)=\infty$ requires $\lam=\infty$. Because of the
rescaling $x(\lam)\rightarrow x(\lam)/\lam^2$ in the vicinity
of $\lam=\infty$ we find $d-2$ solutions, so that
\be
m = 1 + D\cdot b = d-1
\ee
$D$ is isomorphic to the base and so has genus $0$, thus
\be
-2m = \vec{l}^2
\ee
Since (\ref{cubic}) is of degree $6d$ we need to adjust
$(2d+1)+(3d+1)$ parameters  in $x(\lam)$ and $y(\lam)$ to
solve $(6d+1)$ equations in (\ref{cubic}).
This can generically be done only for $d=2$.
For $d=2$ $x(\lam)$ is quadratic and (\ref{cubic}) implies
polynomial equations on the coefficients.
It can be checked that there are exactly 120 solutions for $x(\lam)$.
Taking into account the $\pm$ sign in $y(\lam)$ we find $240$
states. They should combine, as anticipated in \cite{MVII}, with
8 more states to form the $\rep{248}$ of $E_8$.

What are the additional 8 states? They must be in the cohomology
class of $b+f$. There can be no smooth curves in that class
because $(b+f)\cdot f = 1$ implies that it must be a section and thus 
have genus $0$, but the genus formula gives 1 for an irreducible curve
of class $b+f$. Moreover, unlike the case of K3 \cite{BSVtop},
$\SF$ has only one complex structure for a given metric.
Thus the $b+f$ state can only be a bound state of a 2-brane of
class $b$ (i.e. 
the section $x=y=\infty$) with a 2-brane wrapped around $f$.
This can be thought of as a reducible curve $(\inv{x})(z-z_0)=0$.
The fiber has a {\em geometrically} a moduli space of $\CP{1}$
and we should find that there are 8 bound states at threshold.
We will discuss this point in the next section, but let
us go on to larger values of $m$.

States with $m=2$ cannot come from polynomial $x(\lam)$-s
since for $d>2$ there are no generic solutions.
By naive counting of the number of variables (i.e. coefficients in
$x(\lam)$ and $y(\lam)$) we see that for $m=2,3,4$ there can be
 polynomial solutions only for non-generic points in the the moduli
space (coefficients of $f_4$ and $g_6$) and for $m>4$ there are
no polynomial solutions at all.
We will prove  this more rigorously in appendix (B).

The more general meromorphic solution is given by
\be
x(\lam) = {{P(\lam)} \over {Q(\lam)^2}}, \qquad
y(\lam) = {{R(\lam)} \over {Q(\lam)^3}},
\ee
where $P,Q,R$ are polynomials in $\lam$ that satisfy
\be
R(\lam)^2 = P(\lam)^3 - f_4(\lam)P(\lam) Q(\lam)^4 - g_6(\lam) Q(\lam)^6
\label{PQReq1}
\ee
Let
\be
d_Q = \degr{Q(\lam)},\qquad
d_P = \degr{P(\lam)},\qquad
d_R = \degr{R(\lam)}.
\ee
The number of variables (coefficients of $P,Q,R$ up to an overall constant)
is
\be
d_P + d_Q + d_R + 2
\ee
If $d_P > 2 d_Q + 2$ then from (\ref{PQReq1}) $d_R = {3\over 2} d_P$
and the number of equations (coefficients in (\ref{PQReq1})) is
$2 d_R+1 = 3 d_P+1 > d_P + d_Q + d_R + 2$ so generically there
is no solution and in fact (similarly to the proof in appendix (B)),
for $d_P > 2 d_Q + 8$ there is never a solution.
For 
\be
d_P = 2 d_Q + 2,\qquad  d_R = 3 d_Q + 3
\ee
The number of variables equals the number of equations and there
is a finite number of solutions which must correspond to highest weights of
an $E_8$ representation. The momentum $m$ of these states is given
by the intersection with the base $b$, i.e. the number of poles
\be
m = 1 + d_Q
\ee
Since the curve is still isomorphic to the base the genus formula
identifies
the $E_8$ representation as the states with lattice vector $\vec{l}$ 
satisfying
\be
2m = 2 + 2 d_Q = \vec{l}^2
\label{lhighm}
\ee

\subsection{Double covers}

For $n=2$ the curves $(x(\lam),y(\lam))$ intersect
the fiber twice. Thus they are double-valued functions
of $\lam$. They must be of the form
\be
x(\lam) = P(\lam) + Q(\lam)\sqrt{R(\lam)},
\qquad
y(\lam) = S(\lam) + T(\lam)\sqrt{R(\lam)},
\ee
where $P,Q,R,S,T$ are meromorphic functions and the same $\sqrt{R}$
appears in both equations.

The simplest case is
\be
x(\lam) = P_2(\lam),\qquad
y(\lam) = \sqrt{P_2^3 - f_4 P_2 - g_6}
\ee
with
\be
\degr{P_2} \le 2
\ee
This is a double cover of the base with (generically) 6 branch
points. It doesn't intersect $x=y=\infty$ (because of the 
transformation $\xi = {x\over {\lam^2}}$ at the vicinity of infinity).
The genus is $g=2$.
It follows that the cohomology class is
\be
\lbr D \rbr = 2 b + 2 f
\ee
The {\em geometrical}
moduli space of such curves is the coefficients in $P_2(\lam)$.
When any of the coefficients becomes infinite, we can factor
the infinity out to write 
\be
P_2(\lam)\sim \infty\times (\lam-\lam_0)(\lam-\lam_1)
\ee
so it becomes a reducible curve -- a bound state of four
curves: $\lam-\lam_0$, $\lam=\lam_1$ and twice $x=y=\infty$.
We learn from this that the ``direction'' of infinity matters
so that the moduli space of the coefficients of $P_2$
is actually $\CP{3}$.
We note also that at additional 120 points in $\CP{3}$ the curve
becomes reducible into two curves of classes $b+f\pm \vec{l}$
with $\vec{l}^2 = -2$, which are precisely the curves we
discussed in the previous section. This happens when 
$P_2^3 - f_4 P_2 - g_6$ is a perfect square. However, we believe
that again those two curves are in a bound state.

On this $\CP{3}$ we have to fiber the moduli space of flat $U(1)$
bundles on the $g=2$ Riemann surface. This is exactly the
same setting as in \cite{BSVtop} and we find 
$H^*(\SF^3/S_3)$. Indeed as in \cite{BSVtop}, there is a unique
curve in $\SF$ that passes through 3 generic points
in $\SF$.
The conditions for passing through 3 generic points
translates into 3 linear equations in the coefficients of the 
corresponding quadratic polynomial in $\lam$.

%==================================================================%
% Section (5): Reducible curves or bound states of 2-branes        %
%==================================================================%

\section{reducible curves}

We have seen in the previous section that in order to complete
an $E_8$ multiplet there must exist states corresponding to
a 2-brane wrapped on a 2-cycle in the cohomology class $b+f$.
Those states will join the curves in the $b+f+\vec{l}$ class
with $\vec{l}^2 = -2$ and together they will form 248 states
in the $\rep{248}$ of $E_8$.
We saw however, that there is no irreducible curve in the class $b+f$
so the extra required states must be {\em bound states} of two 2-branes
one wrapped on the base $b$ and the other wrapped on the fiber $f$.

The purpose of this section is to argue that such states indeed
exist and their $E_8$ quantum numbers correspond in a natural way
to the Cartan subalgebra of $E_8$ -- that is the 8 missing states in
$\rep{248}$.

The general framework for bond-state questions has been developed
in \cite{WitBST,SenUDU} (and see also \cite{HanKle} for a similar
discussion for 5-branes).

On each of the two 2-branes there live 8 bosonic fields whose zero modes
correspond to translations of the 2-brane in a transverse direction.
One has to determine what low energy fields live on the intersection
and how they interact with the low-energy fields on the two 2-branes.
Then, one has to determine whether the system has a normalizable state
(i.e. its wave-function decays fast enough as the separation between
the 2-branes increases). 

This is very similar to the problem studied in \cite{SenUDU} and indeed
we will use the technology developed there.

Let us start by describing the $E_8$ quantum numbers of such a bound
state.

On a 2-brane of M-theory there lives the dimensional reduction
of $\SUSY{1}$ $U(1)$ SYM from 10D down to 3D, since this
is what lives on a 2-brane of type-IIA \cite{WitBST}.
The 11D Lorenz symmetry is manifest only after dualizing
the gauge field to obtain 8 scalars on the world-volume which
represent transverse oscillations.

For the 2-brane that wraps around the base $b$, 4 out of the 8
scalars are massive, because the base is an isolated 2-cycle
in the CY. For the 2-brane that wraps the fiber $f$, 2 out of the
8 scalars are massive because the 4-cycle $\SF$ is also isolated.
Out of the remaining 6 scalars 2 have the moduli space which is the
base $\CP{1}$ corresponding to moving the fiber along the base.

Although the {\em dynamical} degrees of freedom are the same
after the world-volume  duality that replaces
the gauge field with the 8th scalar,
we do loose the global degrees of freedom of flat gauge connections.
In our case we do have to take into account the flat gauge connections
of the 2-brane that wraps around the fiber $f$.
The resulting setting is exactly as in \cite{BSVtop}.
The moduli space of flat gauge connections on the fiber is isomorphic
to the fiber itself and the entire moduli space is naturally
isomorphic to $\SF$. The states that we obtain correspond to the
cohomology of this moduli space.
Out of $h^{1,1}(\SF) = 10$, 8 states naturally correspond to
the Cartan of $E_8$ and those are the states that complete the
multiplet. In addition there are 4 more neutral states from 
the remaining two in $h^{1,1}$ and from $H^4$ and $H^0$.
They are in the $\rep{1}\oplus\rep{3}$ of $SU(2)_1\subset SO(4)$.

Now we have to address the question of whether the bound state
exists at all.

\subsection{Bound states in type-IIA}

Once we compactify the 11th direction we can study the bound state
question as a quantum-mechanical question.
On the $0+1$ dimensional intersection of the two 2-branes
the massless degrees of freedom are: 4 complex fermions in the 
$\rep{4}$ of $SO(5)$ (the rotations in directions transverse to
the two 2-branes) and two complex bosons which are $SO(5)$ scalars
but are in the $(+\hlf,+\hlf)\oplus(-\hlf,-\hlf)$ of 
$SO(2)\times SO(2)$ (rotations in the 2-brane world-volume keeping
the $0+1$ intersection fixed).
Those states are oppositely charged under the difference of the two $U(1)$-s
on the two 2-branes and neutral with respect to the sum.
The $0+1$ dimensional theory turns out to be the dimensional
reduction of 4D $\SUSY{2}$ $U(1)$ vector-multiplet coupled to a charged 
hyper-multiplet. The vector multiplet is the reduction
of the difference of the (super) $U(1)$-s that live on the 2-branes
and the hyper-multiplet comes from the fields on the intersection.
Since the areas of the 2-branes are finite, the vector-multiplet
is also dynamical. 

Now we have to take account of the fact that 2 out of the 5 transverse
directions are ``frozen'' because those are directions inside
the Calabi-Yau and the 4-cycle $\SF$ where all the 2-branes live
is isolated. 
This amounts to decomposing the fields as 4D $\SUSY{1}$ multiplets
and making the chiral multiplet which is part of the $\SUSY{2}$
vector multiplet massive.
We end up with the dimensional reduction of 4D $\SUSY{1}$
vector multiplet coupled to two oppositely charged chiral
multiplets.

\subsection{Is there binding in the 11th direction as well?}

So far we know that there is stability with respect to separation
in directions $1\dots 4$ when the 11th dimension $x_{10}$
is compactified.
Using the techniques of \cite{SenUDU} we can also determine
whether the state is stable in the 11th direction or not.

As argued in \cite{SenUDU}, if a bound state exists then it is
possible to find a state with $\hlf$ a unit of momentum
around the 11th direction for each 2-brane.
If the bound state did not exist
the minimal momentum around the 11th direction for each 2-brane
would be 1 unit.

The momentum of a 2-brane around the 11th direction is
\be
\int_{\mbox{2-brane}} \dot{\phi}
\ee
 where $\phi$ is the coordinate
along the 11th direction, which is dual on the 2-brane world-volume
to the gauge field $A_\u$.
Thus the momentum is given by  
\be
\int_{\mbox{2-brane}} F_{12}
\ee
where $F_{12}$ is the magnetic field on the 2-brane.
So we are looking for a state with half the Dirac unit of magnetic
flux on each 2-brane. Stated differently, the two 2-branes have a $U(1)$
holonomy of only $\pi$ around the origin (where the other
2-brane intersects it). 

What is the physical manifestation of that?

We recall that the $0+1$ dimensional scalars coming from the DD
states connecting the two 2-branes had spinor quantum numbers
under rotations in the 89 direction or the 67 direction (i.e.
on the 2-brane world-areas). In addition, those scalars were
charged one unit under the $U(1)$ of each 2-brane. 
This means that the field changes sign under a $2\pi$ rotation
in either of the 2-branes. This is no problem if the wave-function
$\Psi$ is even in those scalar variables, but odd wave-functions
can only be defined if there is a $U(1)$ holonomy of $\pi$ 
in each 2-brane, so as to make the field single-valued.

Thus, to determine whether there is binding in the 11th dimension
as well we must find a wave-function that is normalizable 
and {\em odd} with respect to the scalar variables.

We will not go on and explicitly solve the QM problem but
we believe that such a state exists.

%==================================================================%
% Section (6): Physical interpretation                             %
%==================================================================%
\section{Physical interpretation}

Now we can compare the states corresponding to irreducible curves
and bound states of curves to the CFT that lives on the
non-critical string.
Taking $n=1$ in (\ref{dclass}) we have BPS states of the string which
wind once.
The full Fock space is of the form
\be
\prod_l \widetilde{b}^{A_l\a_l}_{-n_l''}
\prod_k \widetilde{a}^{\v_k}_{-n_k'}
\prod_i a^{\u_i}_{-n_i} \prod_j a^{I_j}_{-m_j} |p_L, \theta\ket
\ee
\be
\u_i = 1\dots 4, \qquad A_i = 1,2,\qquad I_i = 1\dots 8
\ee
The bosonic oscillators $a^\u_{-m}$ and $\widetilde{a}^\u_{-m}$
corresponding to transverse oscillations
have indices in the $\rep{4} = (\rep{2},\rep{2})$ of
 $SO(4)=SU(2)_1\times SU(2)_2$ while the fermionic operators
$\widetilde{b}^{A\a}_{-m}$ are in $2(\rep{2},0)$.
$a^\u_{-m}$ are left-moving while $\widetilde{a}^\u_{-m}$ and
$\widetilde{b}^{A\a}_{-m}$ are right movers.
The $a^I_{-m}$ oscillators are left-moving internal $E_8$ lattice
oscillators.

The  states are naturally in the analog of
the Green-Schwarz formalism, but we stress
that we have assumed {\em nothing} about the microscopic
description of the string. All we used is the knowledge of
the low-energy fields that live on the string.

The Green-Schwarz vacuum $|p_L, \theta\ket$
is in the $\rep{3}+\rep{1} + 2(\rep{2})$ of $SU(2)_1$
where $\theta$ specifies the $SO(4)$ state and $p_L$ specifies
the point on the (dual of the) internal $E_8$ lattice.

BPS states have only left-moving excitations.
\be
\prod_i a^{\u_i}_{-n_i} \prod_j a^{I_j}_{-m_j} |p_L, \theta\ket
\label{states}
\ee

Relating (\ref{states}) to (\ref{dclass}) we find
\bear
m &=& (L_0 - \bar{L}_0) = L_0 = \sum n_i + \sum m_j \nn\\
\vec{l} &=& p_L \nn
\eear

The spin of the state in the 4 transverse directions
corresponds to the $SO(4)$ representation
of (\ref{states}) which is deduced from the $\u_i$-s.
In the geometrical picture we quoted from \cite{WitPMF} that 
the $SO(4)=SU(2)_1\times SU(2)_2$ quantum numbers are calculated from
the Lefschetz decomposition of $H^*({\cal M})$ and in particular
the $SU(2)_2$ spin is at most $\hlf$ and the $SU(2)_1$ spin
is at most $\hlf {\mbox{dim}}_\IC {\cal M}$. The $SU(2)_2$ spin
states are related to each other by supersymmetry.
We will see shortly that this procedure has to be modified
because of the special features of our moduli space.

For $m=0$ there was one state $b$ corresponding to $|0\ket$.
For $m=1$ we found $2\times 120$ states from 2-branes
 wrapped on irreducible curves. They were
supplemented with 8 more scalar bound states of $b$ and $f$
to form the $\rep{248}$ of $E_8$. Those states correspond to
$|p_L^2 = -2\ket$ and $a^I_{-1}|0\ket$.
There were 4 additional states from $H^*({\cal M})$
of the bound state. Those have to correspond to $a^\u_{-1}|0\ket$
and thus be in the $(\rep{2},\rep{2})$ of $SO(4)$ 
but according to the Lefschez decomposition they would
have to be in $(\rep{1},\rep{1})\oplus (\rep{3},\rep{1})$
so at first sight we seem to have a discrepancy.

What is the difference between our situation and that of
\cite{WitPMF}? 

In our case the moduli space is that of the bound state of $b$
and $f$. As we saw in section (5), the fermions that live
on the $0+1$ dimensional intersection of $b$ and $f$
are in a non-trivial representation of $SO(4)$. Thus they modify
the $SO(4)$ quantum numbers of the states and eventually we will
obtain $(\rep{2},\rep{2})$ as it should be.

Moving on to higher values of $m$ the geometrical states are either
irreducible curves with $E_8$ representations satisfying (\ref{lhighm})
or a bound state of an irreducible curve with $0\le m'<m$ and 
$t = m - m'$ 2-branes wrapping the fiber $f$.
The moduli space of such a configuration is
\be
{\cal M} = \SF^t / S_t
\ee
where each $\SF$ comes from one fiber and
$S_t$ is the permutation group, since the 2-branes are indistinguishable.
The corresponding states will come from $H^*(\SF^t / S_t)$ for each
curve corresponding to $m'$.

The construction of the cohomology $H^*(\SF^t / S_t)$
was described in \cite{VWSCT}.
One has to construct a Fock space from operators $c_{-n}^w$ where
$w$ corresponds to the states of $H^*(\SF)$.
Thus, for $w=e_1\dots e_8$ we get 8 bosons which comprise
the ``internal'' $E_8$ lattice and would be
in 1-to-1 correspondence with the $a^I_{-n}$ above.
The other 4 values of $w$
are $w=b,f$ and the two extra states $H^0$ and $H^4$.
Those states are in the $\rep{3}+\rep{1}$ of $SU(2)_1\subset SO(4)$
but have to be modified to $(\rep{2},\rep{2})$ because of the fermions at
the intersection of the fiber and the base. They will become
the $a^\u_{-n}$ of (\ref{states}).
The irreducible curve with ``momentum''  $m'$  corresponds
to the point $\vec{p}_L$ in the dual of the internal $E_8$ lattice.
Thus it seems that the geometrical construction indeed reproduces
the Fock space. We have made the assumption that the for each $p_L$
there corresponds exactly one irreducible curve.
In principle, checking this assumption amounts to counting the
number of solutions to polynomial equations in the coefficients
of $P,Q,R$ in (\ref{PQReq1}).
(A straightforward count of the degrees of
the resulting equations, however, didn't seem to work because
the equations turned out to have multiplicities. It is probably
easier to count the curves in another representation of $\SF$, e.g.
a blow-up of $\CP{2}$. It is also interesting to check whether
the reducible curves become irreducible after a deformation
away from the elliptic fibration.)

Next come the double covers with $n=2$.
We will consider only the states in the class $2b+2f$.
Their quantum numbers agree with the interpretation as bound states
of two singly-wound strings or as a single doubly-wound string.
In F-theory a doubly-wound non-critical string has an enhanced
$U(2)$ gauge symmetry on it \cite{WitBST} and the 16 fermions
that generate the $E_8$ are in the $\rep{2}$ of $U(2)$.
In 6D we do not expect a bound state 
(this is $U(2)$ with  $\SUSY{4}$ supersymmetry).
However, after compactification we can, as in \cite{GanHan},
 put an appropriate $SO(16)\times SO(16)$
Wilson line and T-dualize to the $SO(32)$ 
small instanton of \cite{WitSML}.
In that theory we are in the Coulomb phase of $Sp(1)$
\cite{GanHan} and there are the $Sp(1)$ gauge bosons
whose charge corresponds to $n=2$ in (\ref{dclass}).

\section*{Acknowledgments}
I wish to thank V. Balasubramanian, A. Hanany, I.R. Klebanov,
S. Ramgoolam, C. Schmidhuber and E. Sharpe for helpful discussions.

%==================================================================%
% Appendix (A): The geometry of the surface                        %
%==================================================================%
\appendix
\section{On the geometry of $\SF$}
Starting with the elliptic fibration
\be
y^2 = x^3 - f_4(\lam)x - g_6(\lam)
\ee
Which describes a complex surface $\SF$ fibered over the exceptional
divisor.
Let 
\be
\Delta(\lam) = {{f^3}\over {27}} - {{g^2}\over {4}}
=\prod_{i=1}^{12} (\lam - e_i)
\ee
We assume that $\infty$ is not a root.
We want to describe the 10 2-cycles of this surface.
2 of them are the base and fiber. The remaining form an
$E_8$ intersection matrix and project to paths from
$e_i$ to $e_j$ on the base. The condition is that
we can find a 1-cycle on the fiber which degenerates at  $e_i$
and as we continuously move along the path $\gamma_{ij}$ from
$e_i$ to $e_j$, we end up again with a degenerate cycle at $e_j$.
All the 1-cycles will then form a 2-cycle with the topology of 
a sphere.

Let $e_i$ be the vertices of a 12-gon. Let the region inside
the 12-gon be $R$. We can define 
\be
\sqrt{\Delta} = \sqrt{\prod_{i=1}^{12} (\lam - \a_i)}
\ee
to be single-valued inside $R$.
The roots of the cubic
\be
x^3 - f_4(\lam)x -g_6(\lam) = \prod_{k=1}^3 (\lam-\a_i)
\ee
are given by
\be
\a_k = e^{{{2\pi i k}\over 3}} 
\left({g\over 2} + \sqrt{\Delta}\right)^{1/3}
+
{{f/3}\over {
e^{{{2\pi i k}\over 3}}({g\over 2} + \sqrt{\Delta})^{1/3}}}
\ee
When $\Delta = 0$, $\a_1 = \a_2$.
Let us draw cuts for $\sqrt{\Delta}$ from $\a_{2k-1}$
to $\a_{2k}$.
Then there are $\IZ_2$ monodromies around each $\a_{2k}$
but as we shall soon see, there are also $\IZ_3$ monodromies
(in the order $\a_1,\a_2,\a_3$) when making a loop
around each of the cuts.
In fact, $\sqrt{\Delta}$ is defined on a double cover which
is a Riemann surface of genus 5. The monodromies are
around the non-trivial 1-cycles.
The cubic has 3 roots for each $z$.
When going around cycles around the $\a_{k}$-s the 3 roots
get interchanged. The monodromies are thus in $S_3$.
The roots can be ordered so that the
monodromies for loops that start at the origin and go around
the $\a_{k}$-s as follows:
\be
{\mbox{Monodromy }} =
\left\{\begin{array}{ll}
(12) & {\mbox{for a loop around $\a_{2k+1}$}} \\
(13) & {\mbox{for a loop around $\a_{2k+2}$}} \\
\end{array}
\right.
\ee
Let us find 8 2-cycles whose intersection matrix
is $E_8$.
To find the 2-cycles in $H^2(S,\IZ)$ we must pick paths from $\a_k$
to $\a_j$ in such a way that when we start with a 1-cycle in the torus
that shrinks at $\a_k$ and see what happens to the cycle as we move
along the path $\gamma$ to $\a_j$ then it has to shrink again as
we reach $\a_k$.
Every 2-cycle as above has a self-intersection
of $(-2)$.
Cycles that go from $\a_1$ to $\a_{2k+1}$ (like $\a$ in Fig C) from
the outside, have monodromy:
\be
(12)(123)^k (12)(123)^k
\ee
and this is $(1)$ only if $k=3$.
Cycles that go from $\a_1$ to $\a_{2k}$ from the outside have
monodromy
\be
(123)
\ee
and are never trivial.
Cycles that go from $\a_2$ to $\a_{2k+1}$ have also monodromy
$(123)$ and are never trivial.

Now we can give a basis for $H^2(\SF,\IZ)$.
We take the 8 2-cycles: (1) from $\a_1$ to $\a_3$, (2) from $\a_2$
to $\a_{12}$, (3) from $\a_{12}$ to $\a_4$, (4) from $\a_4$ to $\a_6$,
(5) from $\a_5$ to $\a_7$, (6) from $\a_7$ to $\a_9$, (7) from 
$\a_8$ to $\a_{10}$ and (8) from $\a_5$ to $\a_{11}$ but the last one
is from the outside.
\vskip 0.5cm
Their intersection forms the $E_8$ matrix.
%%%%%%%%%%%%
%%%%% Fig.A
%%%%%%%%%%%%
\vskip 0.5cm
\putfig{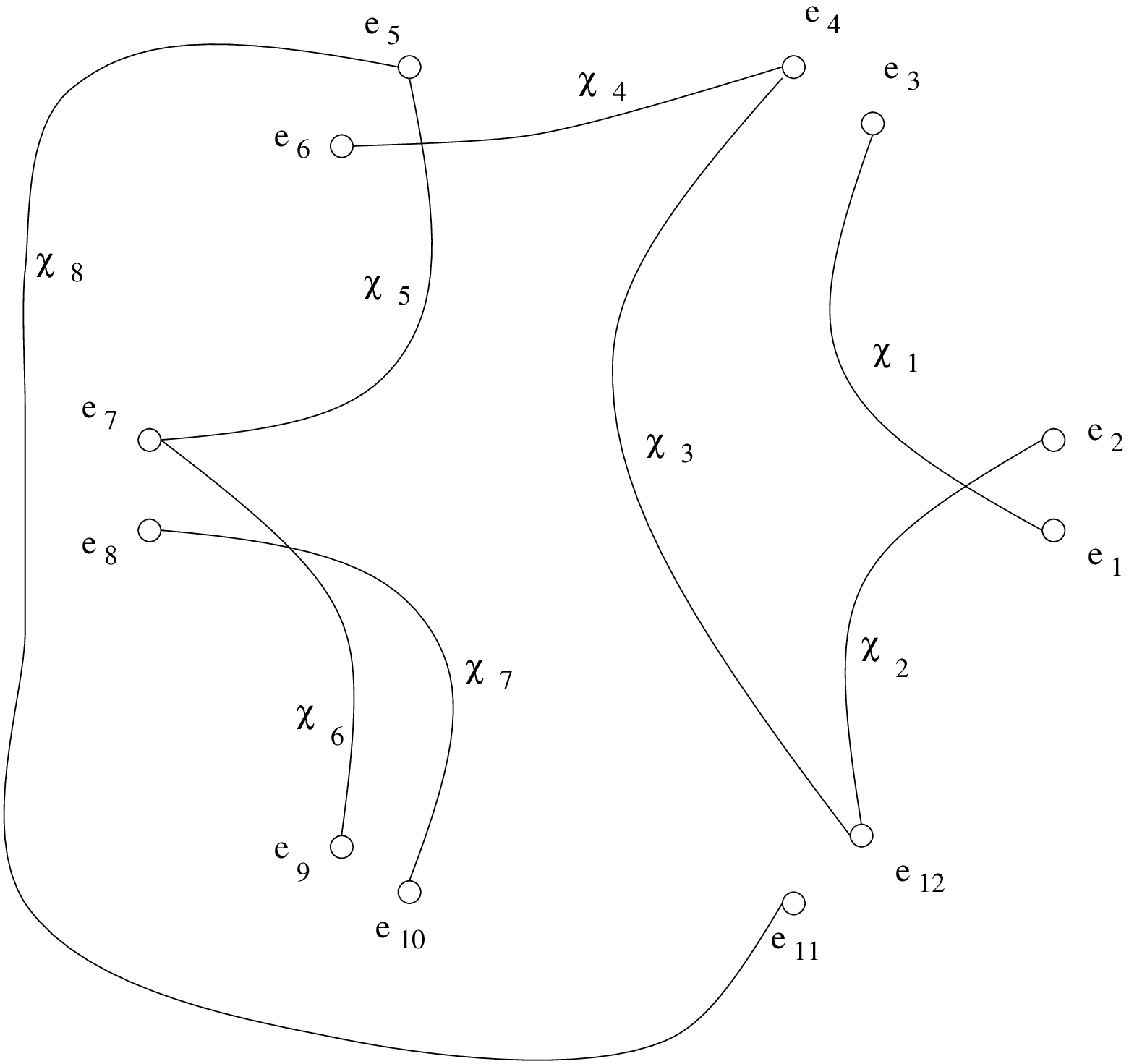}{\it Fig.A: The 2-cycles as paths}{60mm}
\vskip 0.5cm

%==================================================================%
% Appendix (B): On the existence of polynomial sections            %
%==================================================================%
\section{Polynomial sections}

We want to prove that there are no polynomials
$y(\lam)$ and $x(\lam)$ such that 
\be
\degr{x(\lam)}>8
\ee
and
\be
y^2 = x^3 - f_4 x - g_6
\ee
Let $e_i(\lam)$ be the three roots of the cubic.
The functions $e_i(\lam)$ give rise
to a single valued function $e(z)$ that
lives on a triple cover
of the $\lam$-plane which is of genus 4:
\be
2g-2 = 3\cdot(-2) + 12 = 6
\ee
and there are 12 simple branch-points.
Let's call this Riemann surface $\Sigma_4$.
Now
\be
y^2 = (x-e_1)(x-e_2)(x-e_3)
\ee
Each zero of the RHS is at least a double zero.
Now consider the function 
\be
\mu_i =\sqrt{x-e_1}
\ee
Either each zero of $x-e_i$ is a double zero,
in which case $\mu_i$ corresponds to a single-valued
function $\mu(z)$ on $\Sigma_4$ or $x-e_i$ and $x-e_j$ 
have a common root. This can only happen when $e_i=e_j$
i.e. at one of the 12 branch points of the cover
$\Sigma_4\rightarrow \CP{1}$.

Let us start with the first case.
\be
x(\lam(z)) = \mu(z)^2 - e(z)
\label{mue}
\ee
where $\lam(z)$ is a rational function on $\Sigma_4$
with three simple zeroes (and hence three simple poles) and
with 12 simple zeroes of its derivative.
$\mu$ and $e$ and $x(\lam)$ are rational functions on $\Sigma_4$.
Let us count the number of poles of the LHS and RHS.
The poles can be at one of the 3 points $\lam^{-1}(\infty)$.
$e(z)$ has 6 simple zeroes (zeroes of $g_6$) and thus three double
poles. Let $n$ be the degree of $x(\lam)$. Then $x(\lam)$ has
three poles of multiplicity $n$. We see that for $n>2$
$\mu(z)^2$ also has three poles of multiplicity $n$.
Thus $n=2m$ is even and $\mu$ has three poles of multiplicity 
$m$.

Now we proceed as follows.
For a generic $\lam_0\in \CP{1}$ there are three values of $z_i$ with
$i=1,2,3$ for which $\lam(z_i) = \lam_0$. For non-generic $\lam_0$
we take appropriate multiplicities.
Let
\bear
\widetilde{\Delta}(\lam_0) 
&=& \prod_{1\le i< j\le 3} (\mu(z_i)-\mu(z_j))^2 \nn\\
\Xi(\lam_0) 
&=& \prod_{1\le i< j\le 3} (\mu(z_i)+\mu(z_j)) \nn
\eear
$\widetilde{\Delta}$ and $\Xi$ are single valued functions of $\lam_0$
and thus are polynomials.
Furthermore
\be
\widetilde{\Delta} \Xi^2
=\prod_{i< j} (\mu(z_i)^2-\mu(z_j)^2)^2
=\prod_{i< j} (e(z_i)-e(z_j))^2 = \Delta_{12}
\label{dxx}
\ee
where we used (\ref{mue}) and $\Delta$ is the discriminant
\be
\Delta_{12}(\lam) = {{f^3}\over {27}}- {{g^2}\over {4}}
\ee
Let us calculate the degrees of the polynomials $\Xi$ and $\widetilde{\Delta}$.
Near $\lam=\infty$ and for $n>2$ 
\be
\mu_i = \pm\sqrt{x(\lam) + e_i} \sim \pm \lam^m
\label{signs}
\ee
The $\pm$ signs for the three $\pm_i$-s can either all be the same
or be two of one kind and the third of the opposite sign.
In the first case
\be
\Xi = \prod_{i< j} (\mu(z_i)+\mu(z_j)) \sim \lam^{3m}
\ee
but then $\Xi^2$ has degree $6m$ which is greater than $12$
for $m>2$ and contradicts (\ref{dxx}).
for $m=2$ we find 
\be
\widetilde{\Delta} = \const
\ee
we will soon discuss this case as well, but let us first 
see what happens when the signs  in (\ref{signs}) are different,
say two $(+)$'s for $\mu_1$ and $\mu_2$ and one $(-)$ for $\mu_3$.
Then
\bear
\mu_1 + \mu_2 &=& \sqrt{x +e_1} + \sqrt{x+e_2}\sim \lam^m \nn\\
\mu_1 + \mu_3 &=& \sqrt{x +e_1} - \sqrt{x+e_3}
= {{e_1 - e_3}\over {\sqrt{x +e_1} + \sqrt{x+e_3}}}\sim \lam^{2-m} \nn\\
\mu_2 + \mu_3 &=& \sqrt{x +e_2} - \sqrt{x+e_3}\sim \lam^{2-m} \nn\\
\Xi &\sim& \lam^{4-m} \nn
\eear
we see that (\ref{dxx}) requires $m\le 4$.
The remaining cases have
\be
\degr{\Xi} \le 3
\ee
Now we can write define the symmetric sums
\bear
\s_1 &=& \sum\mu_i \nn\\
\s_2 &=& \sum\mu_i^2 = 3x(\lam) + \sum e_i = 3x(\lam) \nn\\
\s_3 &=& \sum\mu_i^3\nn\\
\s_4 &=& \sum\mu_i^2 = \sum(x(\lam) + e_i)^2 = 3x(\lam)^2 + 2f_4(\lam) \nn
\eear
some algebra gives:
\bear
\Xi &=& \prod_{i< j} (\mu(z_i)+\mu(z_j))
=\inv{3} (\s_1^3 - \s_3) \nn\\
0 &=& 6\s_4 + 6\s_2\s_1^2 - 3\s_2^2 - \s_1^4 - 8\s_1\s_3 \nn
\eear
so we find
\be
(\s_1^2 - x)^2 = {4\over 3}f_4 - {8\over 3}\Xi \s_1
\ee
The RHS is a polynomial of degree 4 at most.
Thus $\s_1^2 - x$ is of degree 2 and we find
\be
x = p_m^2 + q_2
\ee
where $q_2$ is of degree 2 (at most) and $p_m$ is of degree $m$ (exactly).
Putting back in the original equation we find
\be
y_{3m}^2 = x_{2m}^3 - f_4 x_{2m} - g_6
 =(p_m^3 + {3\over 2}p_m q_2)^2 
-{9\over 4}p_m^2 (q_2^2- {4\over 3}q_2+{4\over 9}f_4)
+ (q_2^3 - f_4 q_2 - g_6)
\ee
we are looking for non-generic solutions, i.e. solutions
for special values of $f_4$ and $g_6$.
Now let's separate into cases:

\subsection*{$m=2$}
This case admits solutions for a co-dimension 1 subspace of the 
parameter space (of coefficients of $f_4$ and $g_6$).
We will see that generically this subspace of moduli space
corresponds to a nonsingular surface.

This case also includes the missing case above where all the signs
of (\ref{signs}) are equal.

In this case every $x_4$ can be written as
\be
x = p_2^2 + q_1
\ee
After some algebra we find the general solution
\bear
f_4 &=& -2 p_2 r_2 + t_1 \nn\\
g_6 &=& -{9\over 4}p_2^2 q_1^2 
-p_2 r_2 q_1 - p_2^2 t_1  - r_2^2 - t_1 q_1 \nn
\eear
Every choice of $p_2$, $q_2$, $r_2$, $q_1$ and $t_1$
(the subscripts are degrees of the polynomials)
gives appropriate $g_6$ and $f_4$.

\subsection*{$m=3$}
Similarly we find
\be
x = p_3^2 + q_2
\ee
and
\bear
f_4 &=& -{9\over 4}q_2^2 + 3 q_2 - v_1 p_3 - c_0 \nn\\
g_6 &=& {{13}\over 4}q_2^3 - 3 q_2^2 +v_1 p_3 q_2  + c_0 q_2
     - v_1^2 - p_3 q_2 v_1 + c_0 p_3^2 \nn
\eear

\subsection*{$m=4$}
\be
x = p_4^2 + q_2
\ee
and
\bear
f_4 &=& -{9\over 4} q_2^2 +3 q_2 - 2 c_0 p_4 \nn\\
g_6 &=& {{13}\over 4} q_2^3 - 3 q_2^2 
     + 2 c_0 q_2 p_4 - 3 p_4 c_0 - c_0^2 \nn
\eear

\end{document}